\documentclass[%
 reprint,
 amsmath,amssymb,
 aps,
]{revtex4-2}

\usepackage{graphicx}% Include figure files
\usepackage{dcolumn}% Align table columns on decimal point
\usepackage{bm}% bold math
\usepackage{mathtools}% For using dcases
\usepackage{multirow}
\usepackage{makecell}

\begin{document}

\preprint{APS/123-QED}

\title{Crystal Nucleation of Highly-Screened Charged Colloids}% Force line breaks with \\

\author{Marjolein de Jager}
 \email{m.e.dejager@uu.nl}
 %\affiliation{Soft Condensed Matter, Debye Institute of Nanomaterials Science, Utrecht University, Utrecht, Netherlands}
\author{Laura Filion}
\affiliation{Soft Condensed Matter, Debye Institute of Nanomaterials Science, Utrecht University, Utrecht, Netherlands}

\date{\today}% It is always \today, today,

\begin{abstract}
We study the nucleation of nearly-hard charged colloidal particles. We use Monte Carlo simulations in combination with free-energy calculations to accurately predict the phase diagrams of these particles and map them via the freezing density to hard spheres, then we use umbrella sampling to explore the nucleation process. Surprisingly, we find that even very small amounts of charge can have a significant effect on the phase behavior. Specifically, we find that phase boundaries and nucleation barriers are mostly dependent on the Debye screening length, and that even screening lengths as small as 2\% of the particle diameter are sufficient to show marked differences in both. This work demonstrates clearly that even mildly charged colloids are not effectively hard spheres.
\end{abstract}

\maketitle

%\tableofcontents

%%%%%%%%%%%%%%%%%%%%%%%%%%%%%%%%%%%%%%%%%%%%%%%%%%%%%
%%%%%%%%%%%%%%%%%%%%%%%%%%%%%%%%%%%%%%%%%%%%%%%%%%%%%

%% Other effects on (nearly-) HS nucleation:
%- \cite{fiorucci2020effect}: Look at the effect of hydrodynamics on the nucleation rate of nearly-hard spheres (WCA with $\beta\epsilon=40$).
%- \cite{zaccarelli2009crystallization}: Look at the effect of polydispersity on the crystallization of hard-sphere glasses.
%- \cite{pusey2009hard}: They studied crystallization and glass formation of hard spheres over a wide range of polydispersity and volume fraction using molecular dynamics (event driven) simulations. They find that increasing the polydispersity slows down crystal nucleation, because the polydispersity reduces the supersaturation.

Hard spheres are, arguably, the archetypical model system for studying colloidal self-assembly and have been instrumental in our understanding of e.g. phase transitions \cite{harland1997crystallization,gasser2001real,auer2001prediction,cacciuto2004onset,auer2003line,iacopini2009crystallization}, glassy behavior \cite{gordon1976hard,speedy1998hard,zaccarelli2015polydispersity}, and defects \cite{bennett1971studies,pusey1989structure,pronk2001point,pronk1999can,van2017diffusion}. 
Experimentally, hard spheres are typically realized as colloidal particles suspended in a solvent. However, in such systems the colloids nearly always carry surface charges and are generally decorated with surface ligands.  This raises an important fundamental question: when can colloids be seen as hard spheres?

This question is particularly crucial when addressing crystal nucleation of hard spheres where a long standing issue is the large discrepancy between predicted and experimentally observed nucleation rates \cite{harland1997crystallization,sinn2001solidification,schatzel1993density,gasser2001real,iacopini2009crystallization,auer2001prediction,auer2004numerical,zaccarelli2009crystallization,filion2010crystal,schilling2011crystallization,wohler2022hard}. 
Since any induced repulsion between the particles will make the particles act as if they are slightly larger, a quantitative comparison to the true model requires a way to assign an effective packing fraction to the charged systems. The most common mapping involves assigning an effective diameter to the particle such that the freezing point matches that of hard spheres \cite{pusey1987observation,pusey1989structure,underwood1994sterically,harland1997crystallization,schatzel1993density,palberg1999crystallization,royall2013search,kawasaki2010formation,filion2011simulation,russo2013interplay,fiorucci2020effect}. Within this mapping, it has been shown that for sufficiently large colloids, steric interactions due to surface ligands can indeed be accurately taken into account \cite{auer2003phase} -- and this also holds for a simple short ranged repulsion like the Weeks Chandler Andersen (WCA) potential \cite{filion2011simulation}. It would be tempting to then assume that this holds for screened charged interactions as well. Here, however, we will show that screened charged interaction clearly deviate from this pattern. We find that only extremely strongly-screened charged interactions -- ones where the screening length is at most a percent of the particle diameter -- can be well approximated as hard sphere when it comes to nucleation.

We consider a system of $N$ electrically like-charged hard spheres of diameter $\sigma$ suspended in a solvent containing ions characterized by an inverse Debye screening length $\kappa$ and Bjerrum length $\lambda_B$. 
Within  Dejaguin-Landau-Verwey-Overbeek (DLVO) theory, the effective interaction potential is given by
\begin{equation}
	\label{eq:pothcyuk}
	\beta\phi(r) = \begin{dcases}
		\beta\epsilon \; \frac{e^{-\kappa\sigma (r/\sigma -1)} }{r/\sigma}  & \quad \text{for } r\geq\sigma,\\
		\infty       & \quad \text{for } r<\sigma,
	\end{dcases}
\end{equation}
with contact value $\beta\epsilon = Z^2\lambda_B/\sigma(1+\kappa\sigma/2)^2$, where $Z$ is the charge of the colloids in electron charge, and $\beta=1/k_BT$, with $k_B$ the Boltzmann constant and $T$ the temperature. 
Note that in the limit of zero charge ($Z\to0$) or infinite screening ($\kappa\sigma\to\infty$), the repulsive hard-core Yukawa potential of Eq. \eqref{eq:pothcyuk} reduces to the hard-sphere potential.
% \begin{equation}
% 	\label{eq:poths}
% 	\beta\phi(r) = \begin{cases}
% 		0  & \quad \text{for } r\geq\sigma,\\
% 		\infty       & \quad \text{for } r<\sigma.
% 	\end{cases}
% \end{equation}
This system has been extensively studied using experiments, simulations, and theory, and the bulk phase behavior is extremely well understood \cite{alexander1984charge, kremer1986phase, robbins1988phase, monovoukas1989experimental, sirota1989complete, hamaguchi1997triple, hynninen2003phase, yethiraj2003colloidal, royall2003new, hsu2005charge, royall2006re, el2011measuring, smallenburg2011phase, kanai2015crystallization, arai2017surface, chaudhuri2017triple,van2013sterically, kodger2015precise}.
% , with impressive quantitative comparisons between theory and experiment \cite{monovoukas1989experimental, sirota1989complete, kanai2015crystallization,royall2003new}. 
Here we focus on systems with short screening lengths, i.e. ranging from 1-4\% of the particle diameter -- similar to those often considered as hard experimentally \cite{pusey1986phase,pusey1987observation,pusey1989structure,harland1997crystallization,palberg1999crystallization,yethiraj2003colloidal,royall2003new,leunissen2007manipulating}. For the contact values, we explore mainly $\beta\epsilon=8,20,39$, and 81, which allows us to compare to the phase diagrams of Ref. \onlinecite{hynninen2003phase}.
An overview of the potentials is shown in Fig. \ref{fig:potentials}a). Additionally, in Fig. \ref{fig:potentials}a) we display the WCA potential with $\beta\epsilon=40$, which was previously shown to accurately replicate the nucleation behavior of hard spheres \cite{filion2011simulation}.

% In Ref. \onlinecite{hynninen2003phase}, Hynninen \textit{et al.} used free-energy calculations to compute the phase diagram of hard-core Yukawa particles for contact values $\beta\epsilon=8,20,39$, and 81. 
% These show that the hard-core Yukawa system forms a face-centered cubic (FCC) crystal when screening is strong, and a body-centered cubic (BCC) crystal when screening is weak. 

% In order to decide on a range for the system parameters, we look at the experimental values for the range of the nearly-hard repulsion. 

% The PMMA particles used in experiments\cite{pusey1986phase,pusey1987observation,pusey1989structure,harland1997crystallization,palberg1999crystallization} have radii ranging from 305 nm to 435 nm all with an additional stabilizer layer of $\sim$10 nm. The shortest possible screening lengths reported for these particles range from $1/\kappa\sigma=0.015$\cite{yethiraj2003colloidal} to $1/\kappa\sigma=0.047$\cite{royall2003new}.
% Thus, we take four values for the screening length, ranging from extremely short ranged ($1/\kappa\sigma=0.01$) to slightly softer ($1/\kappa\sigma=0.04$). For the contact value, we take the same values as used for the phase diagrams in Ref. \onlinecite{hynninen2003phase}, %i.e. $\beta\epsilon=8,20,39$, and 81, 
% such that we can check the freezing and melting packing fractions obtained from our free-energy calculations.  

\newcommand{\figwidthA}{0.935\linewidth}
\begin{figure}[t]
    \begin{tabular}{l}
     a)  \\[-0.45cm]
      \includegraphics[width=\figwidthA]{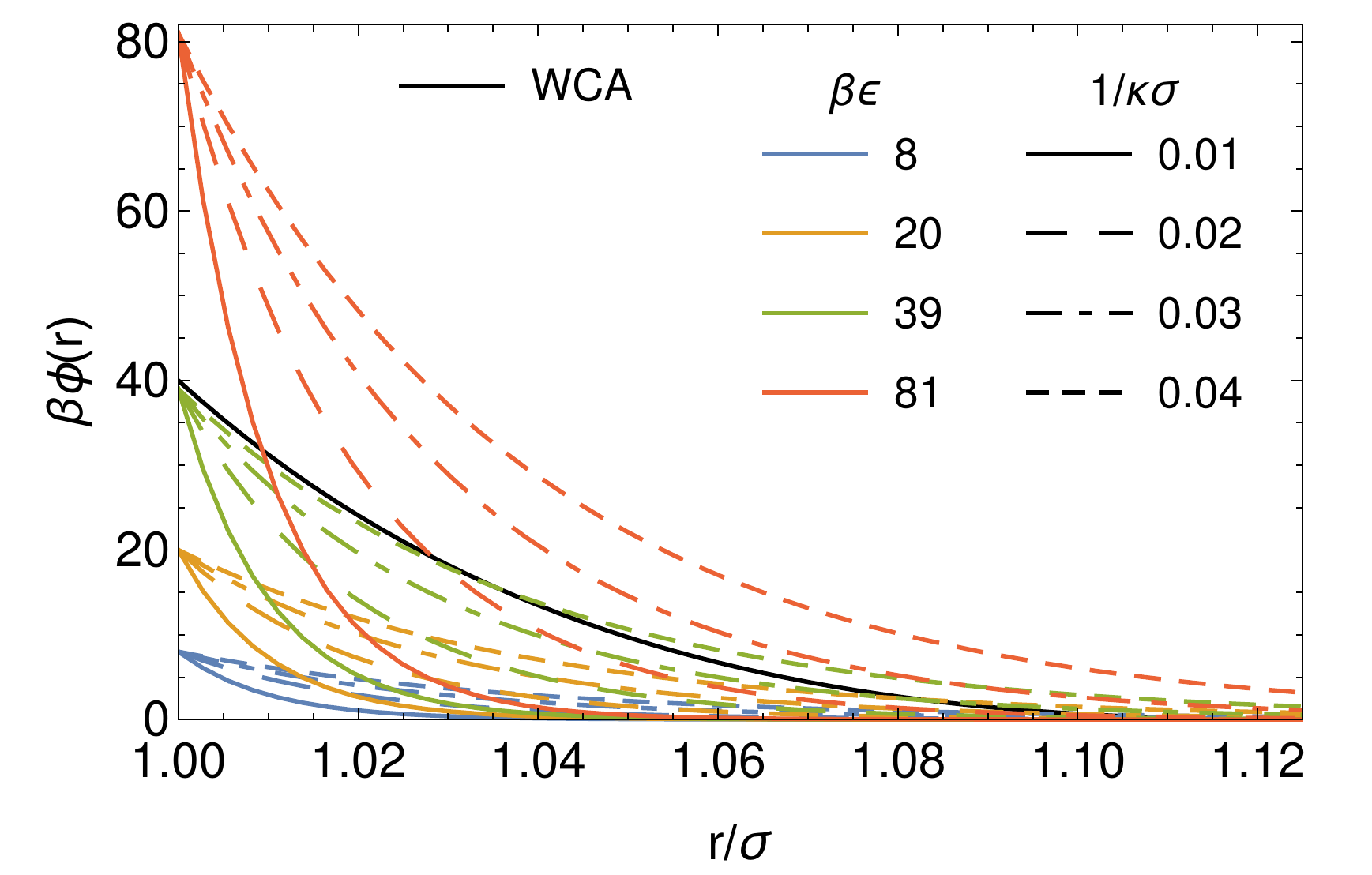} \\
     b)   \\[-0.4cm]
     \includegraphics[width=\figwidthA]{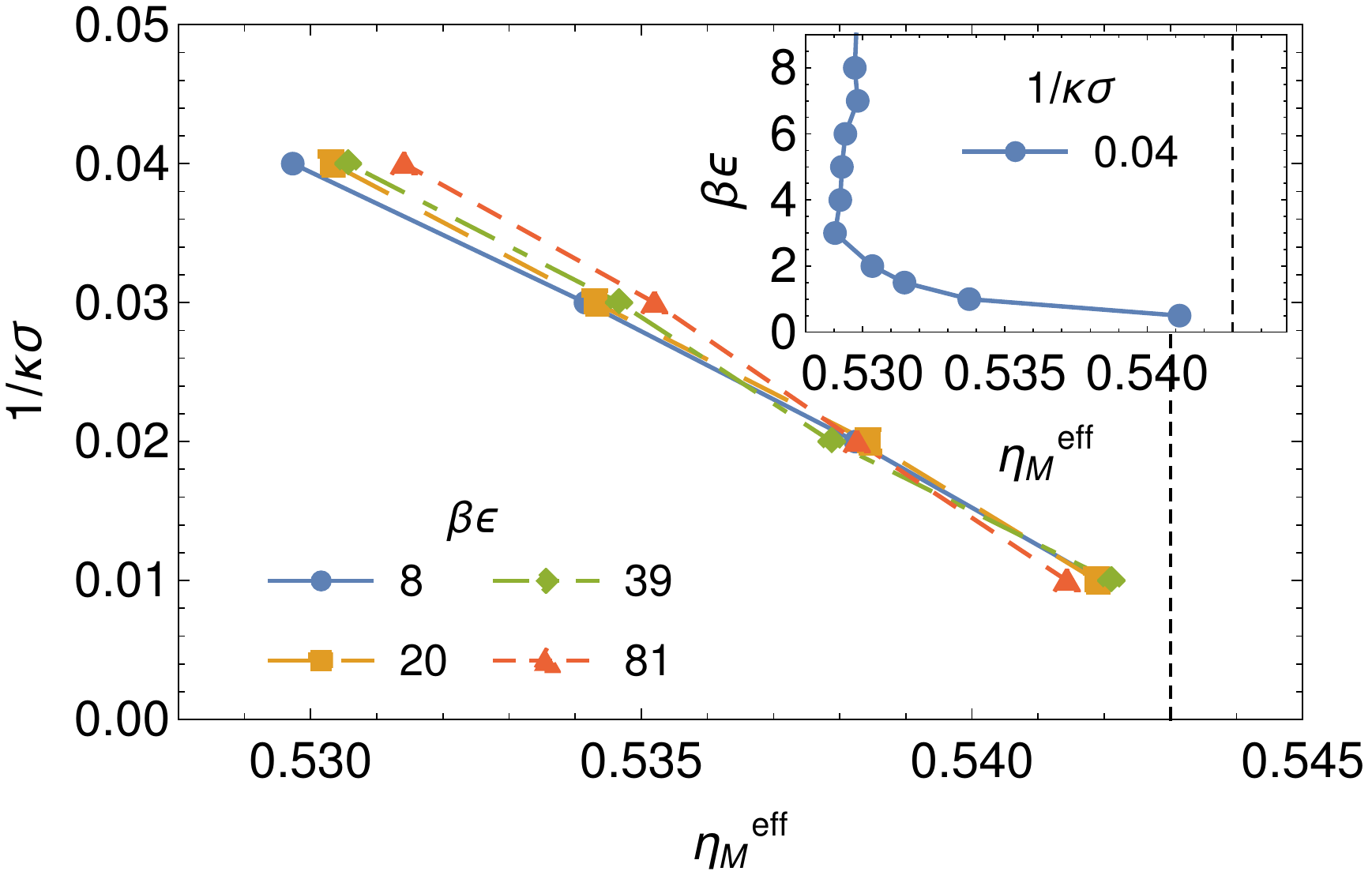}
    \end{tabular}
    \caption{\label{fig:potentials} a) Repulsive tail of the hard-core Yukawa potential for nearly-hard spheres of diameter $\sigma$ with contact value $\beta\epsilon$ (colors) and screening length $1/\kappa\sigma$ (dashing). The solid black line shows the WCA potential with $\beta\epsilon=40$. 
    b) The corresponding effective hard-sphere melting packing fraction $\eta^\text{eff}_M =  (\eta^\text{HS}_F / \eta_F) \, \eta_M$, as a function of $1/\kappa\sigma$. Additionally, the inset gives $\eta^\text{eff}_M$ as a function of $\beta\epsilon$ for $1/\kappa\sigma=0.04$. The vertical dashed lines indicate $\eta^\text{HS}_M=0.543(1)$ \cite{frenkelbook,polson2000finite}. The error in $\eta^\text{eff}_M$ is approximately 0.001.
    }
\end{figure}

We start by mapping the highly-screened hard-core Yukawa particles to hard spheres using an effective hard-sphere diameter via  $\eta^\text{eff} =  (\eta^\text{HS}_F / \eta_F) \, \eta$, where $\eta$ and $\eta_F$ are the actual packing fraction and freezing packing fraction of the system, and $\eta^\text{HS}_F=0.492(1)$ \cite{frenkelbook,polson2000finite} is the freezing packing fraction of hard spheres. %, such that the nearly-hard particles have an effective hard-spheres diameter $\sigma_\text{eff} /\sigma =  (\eta^\text{HS}_F / \eta_F)^{1/3}$.
%, where $\eta_F$ is the freezing packing fraction of the system and $\eta^\text{HS}_F=0.492(1)$ \cite{frenkelbook,polson2000finite} is the freezing packing fraction of hard spheres,  such that one then compares systems at equal effective packing fraction $\eta^\text{eff} =  (\eta^\text{HS}_F / \eta_F) \, \eta$, where $\eta$ is the actual packing fraction of the system,   
To calculate $\eta_F$,  we use free-energy calculations in combination with common-tangent constructions \cite{frenkelbook}. 
%We first compute the equation of state of the fluid and solid phase using both Monte Carlo (MC) simulations in the $NPT$-ensemble and MC simulations in the $NVT$-ensemble combined with the virial equation, with $N\sim 2000$ particles \footnote{We use both methods to increase the accuracy of the equation of state. However, note that we only use the virial equation when the system does not ``feel'' the hard cores. In practise this means that we only use the virial equation for systems with $\beta\epsilon\geq8$.}. 
We first compute the equation of state of the fluid and solid phase using Monte Carlo (MC) simulations in the $NPT$-ensemble with $N\sim 2000$ particles. Additionally, to increase the accuracy of the equation of state, we use MC simulations in the $NVT$-ensemble combined with the virial equation, when the system does not ``feel'' the hard cores. In practice this means that we only use the $NVT$-ensemble combined with the virial equation for systems with $\beta\epsilon\geq8$.
Next, we use thermodynamic integration of the equation of state to get the free energy as a function of the density \cite{frenkelbook}. For the fluid phase, we use the ideal gas as a reference system. For the solid phase, we integrate from a reference density for which we computed the free energy using Einstein integration with finite-size corrections \cite{frenkel1984new,polson2000finite,frenkelbook}. 
A complete summary of all freezing and melting densities is given in the Supplementary Material (SM) and the resulting effective melting packing fractions $\eta^\text{eff}_M$ are shown Fig. \ref{fig:potentials}b). As expected, in the limit of zero charge ($\beta \epsilon \to 0$) the melting density matches that of hard spheres. Surprisingly, however, this limit really requires nearly zero charge, with significant deviations appearing for contact values as low as $\beta \epsilon = 1.$ For $\beta \epsilon>1.5$,  $\eta^\text{eff}_M$ is nearly unaffected by changing $\beta\epsilon$.
As required, in the limit of infinite screening ($1/\kappa\sigma\to 0$) the system reduces to hard spheres. However, as the screening length increases $\eta^\text{eff}_M$ steadily decreases. In the end, only for the strongest screening ($1/\kappa\sigma=0.01$) or smallest contact value ($\beta\epsilon=0.5$) is $\eta^\text{eff}_M$ comparable to $\eta^\text{HS}_M=0.543(1)$ \cite{frenkelbook,polson2000finite}. These results are particularly surprising when looking at a plot of the potentials (Fig. \ref{fig:potentials}a) as it is unclear from their appearance which ones are ``hard enough'' to map well onto hard spheres.  Clearly whether a specific potential will or will not map onto that of hard spheres cannot be seen simply by looking at the potential. 

% An interesting question is how these melting densities compare to previous studies.  For instance, where would the previous studies on very short ranged particles land?  In Fig \ref{fig:potentials}  we also plot the WCA potential and pinpoint its effective melting density. Intriguingly, at first glance the potential appears most like $\beta \epsilon =39$ with a screening length of $0.04$, but the melting density indeed appears to be nearly hard sphere. This can be explained by looking at the tail of the potential (see SI),

% Furthermore, notice that the nearly-hard WCA system maps  $\eta^\text{eff}_M$ well, whereas $\eta^\text{eff}_M$ of the experimental PMMA particles of Pusey and van Megen is comparable to those of the Yukawa particles with $1/\kappa\sigma=0.03$.

%2048 particles for eq of state, finite size with 500 - 2048

%\cite{pusey1986phase} reported $\eta_F=0.407$ and $\eta_M=0.442$ such that $\eta^\text{eff}_M=0.534$ for their PMMA cores of 305 nm with a stabilizer layer of $\sim$10 nm. %Note that the effective hard-sphere diameter is $\sim$20 nm bigger than the PMMA core.

\newcommand{\figwidthB}{0.45\linewidth}
\begin{figure*}[t!]
\begin{tabular}{lll}
     a) & \hspace{0.5cm} & b)  \\[-0.4cm]
     \includegraphics[width=\figwidthB]{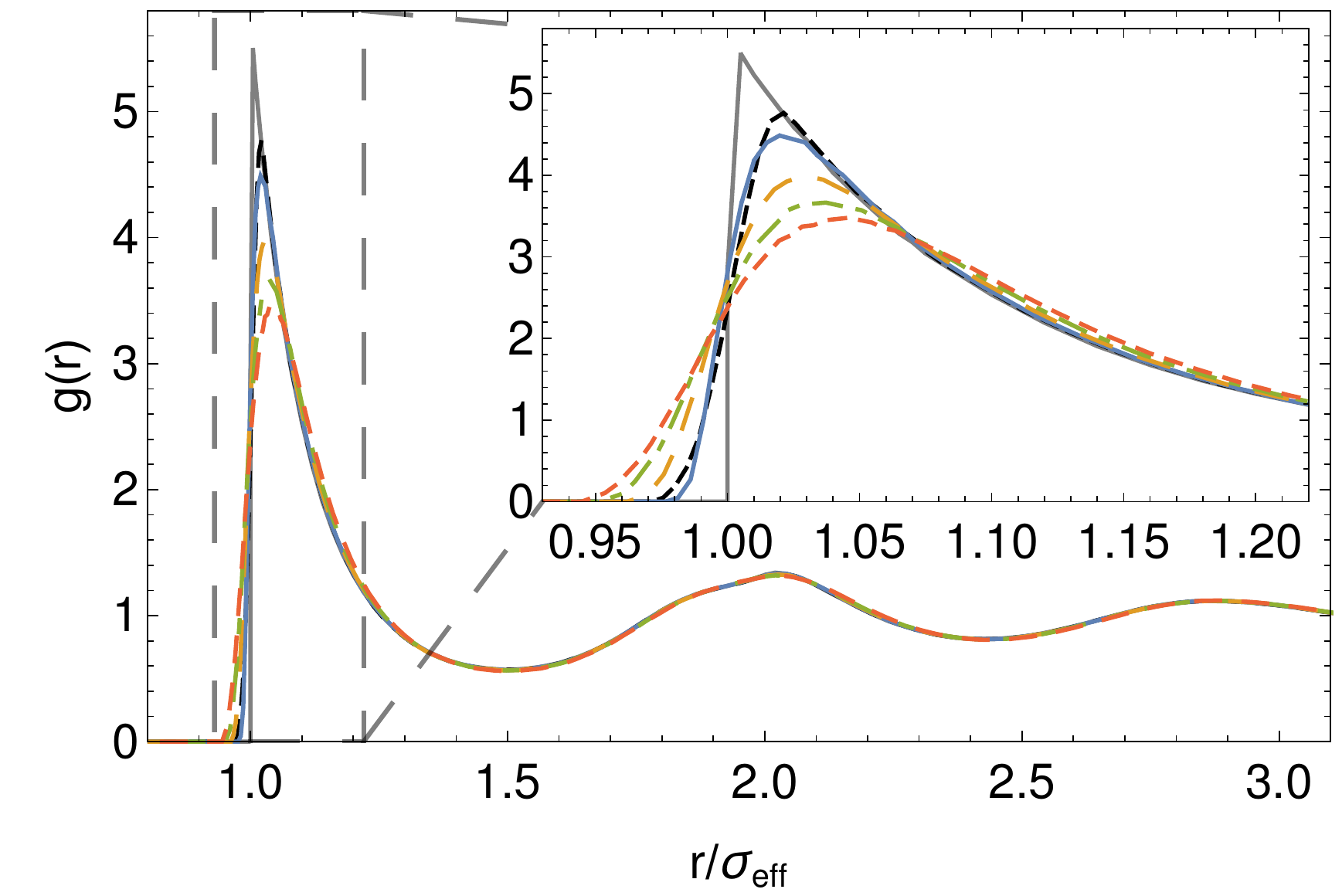} & & \includegraphics[width=\figwidthB]{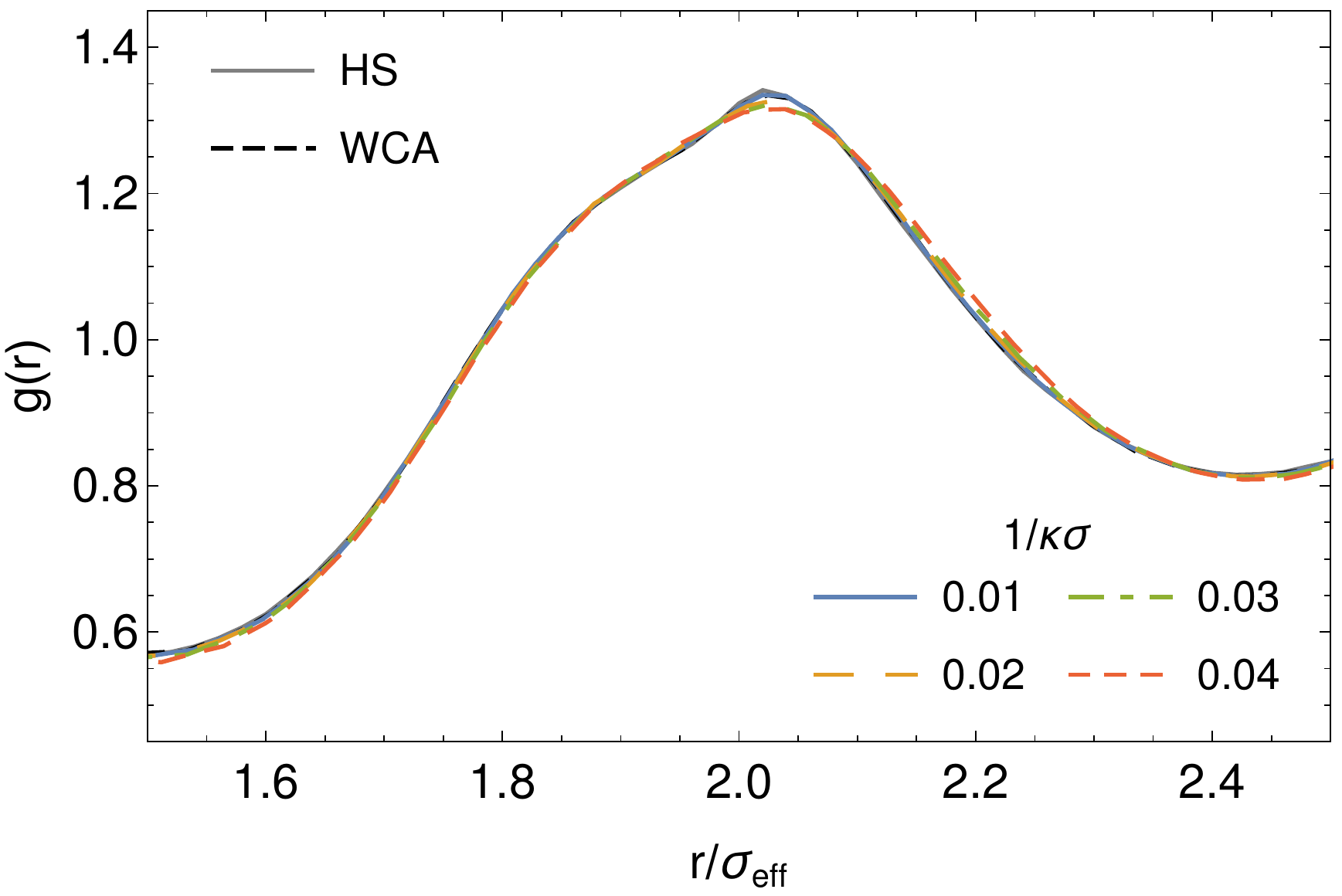} \\
     c) & \hspace{0.5cm} & d)  \\[-0.4cm]
     \includegraphics[width=\figwidthB]{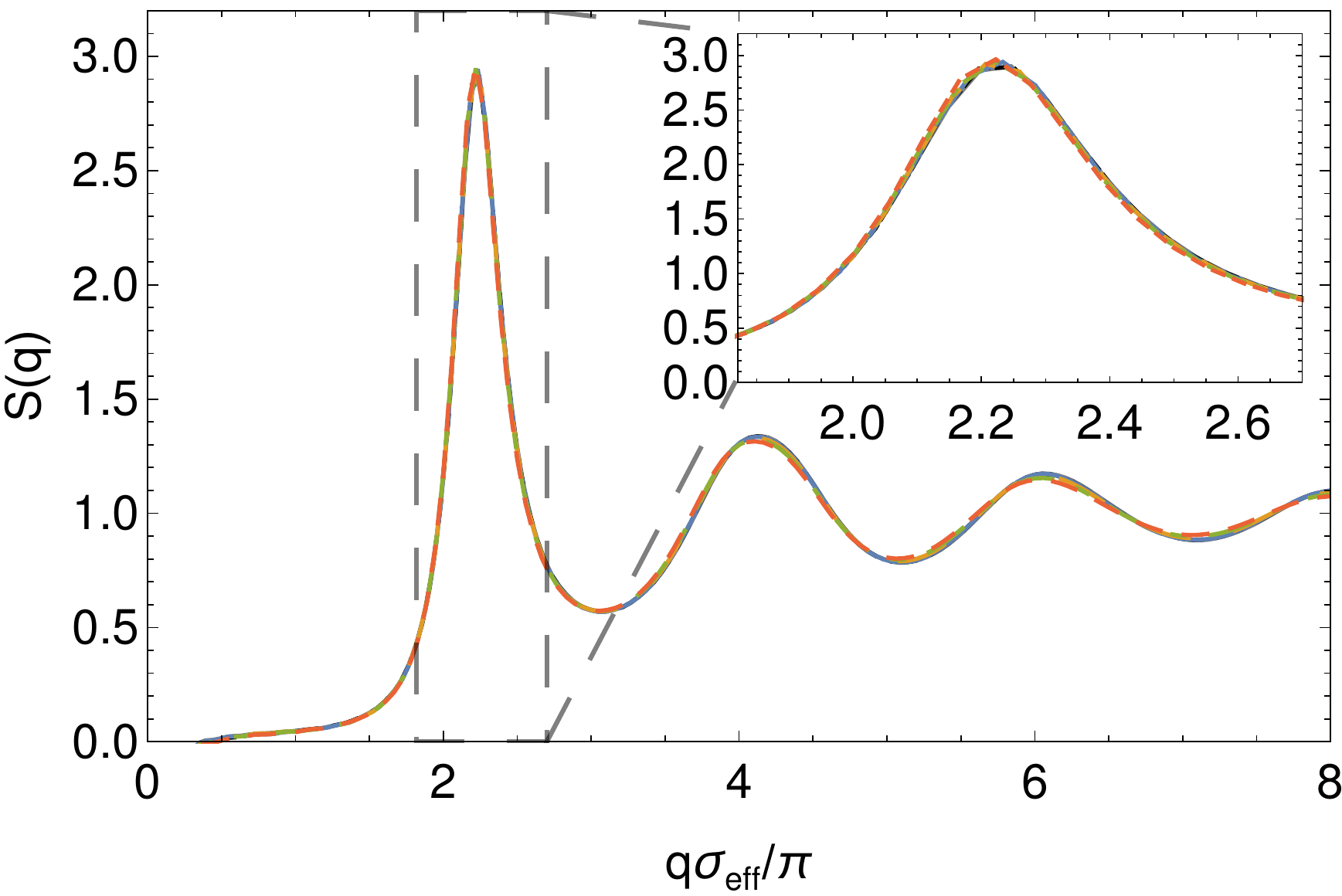} & & \includegraphics[width=\figwidthB]{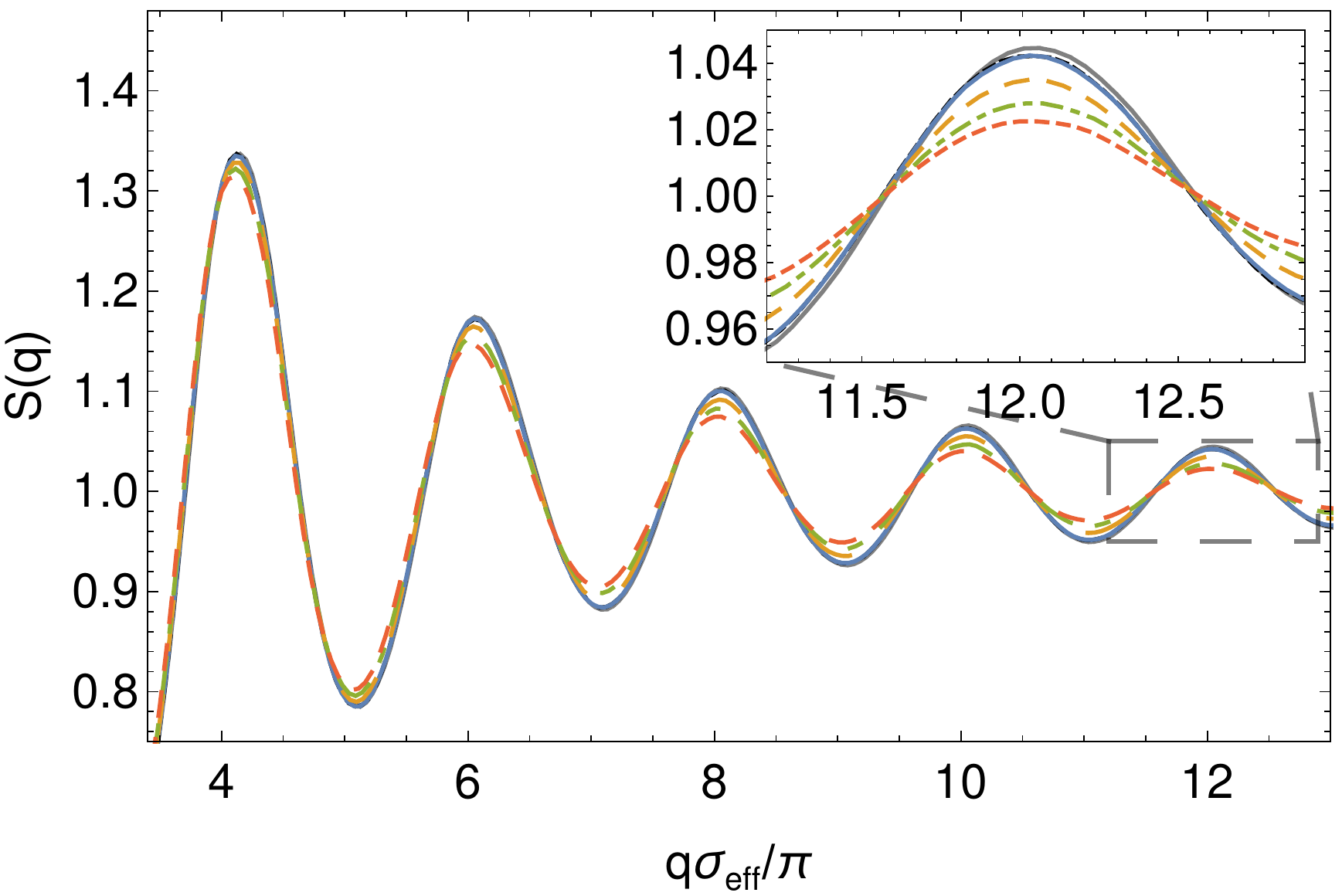}
\end{tabular}
    \caption[width=1\linewidth]{\label{fig:structure} a,b) The radial distribution function $g(r)$ and c,d) structure factor $S(q)$ for a fluid of hard spheres (gray, solid), WCA particles with $\beta\epsilon=40$ (black, dashed), and hard-core Yukawa particles with $\beta\epsilon=39$ and varying screening length $1/\kappa\sigma$ (colors, dashing). All fluids were studied at the freezing packing fraction and the horizontal axes are scaled by the effective hard-sphere diameter $\sigma_\text{eff}$, see SM. 
    The inset of a) zooms in on the first peak of the $g(r)$, and b) shows the second peak. The inset of c) zooms in on the first peak of the $S(q)$, and d) shows the second to sixth peak with the inset zooming in on the sixth peak.
    }
\end{figure*}

We now explore how well these potentials match the structure of the hard-sphere fluid, when mapped to the effective diameter $\sigma_\text{eff} /\sigma =  (\eta^\text{HS}_F / \eta_F)^{1/3}$. 
As the quality of the mapping depends mostly on $1/\kappa \sigma$, here we focus on the structure of hard-core Yukawa particles by varying  $1/\kappa\sigma$  while keeping  $\beta\epsilon=39$ fixed.
%We do this by mapping the radial distribution function $g(r)$ and the structure factor $S(q)$ 
Figure \ref{fig:structure} shows the radial distribution function $g(r)$ and the structure factor $S(q)$ for these fluids at the freezing packing fraction. Notice that all horizontal axes are scaled by the effective hard-sphere diameter, and that this results in an excellent mapping of the $g(r)$ and $S(q)$. The most significant differences between the structure of the hard spheres and the nearly-hard spheres are the very slight broadening of the first peak of the $g(r)$ and the mild shrinking of the higher order peaks of the $S(q)$. These differences increase with increasing softness ($1/\kappa \sigma$), essentially not existing when  $1/\kappa\sigma=0.01$.

%%%%%%%%%%%%%%%%%%%%%%%%%%%%%%%%%%%%%%%%%%%%%%%%%%%%%
%%%%%%%%%%%%%%%%%%%%%%%%%%%%%%%%%%%%%%%%%%%%%%%%%%%%%

%\section{Crystal Nucleation}

%\newcommand{\figwidthA}{0.935\linewidth}
\begin{figure}[h!]
    \centering
    \includegraphics[width=\figwidthA]{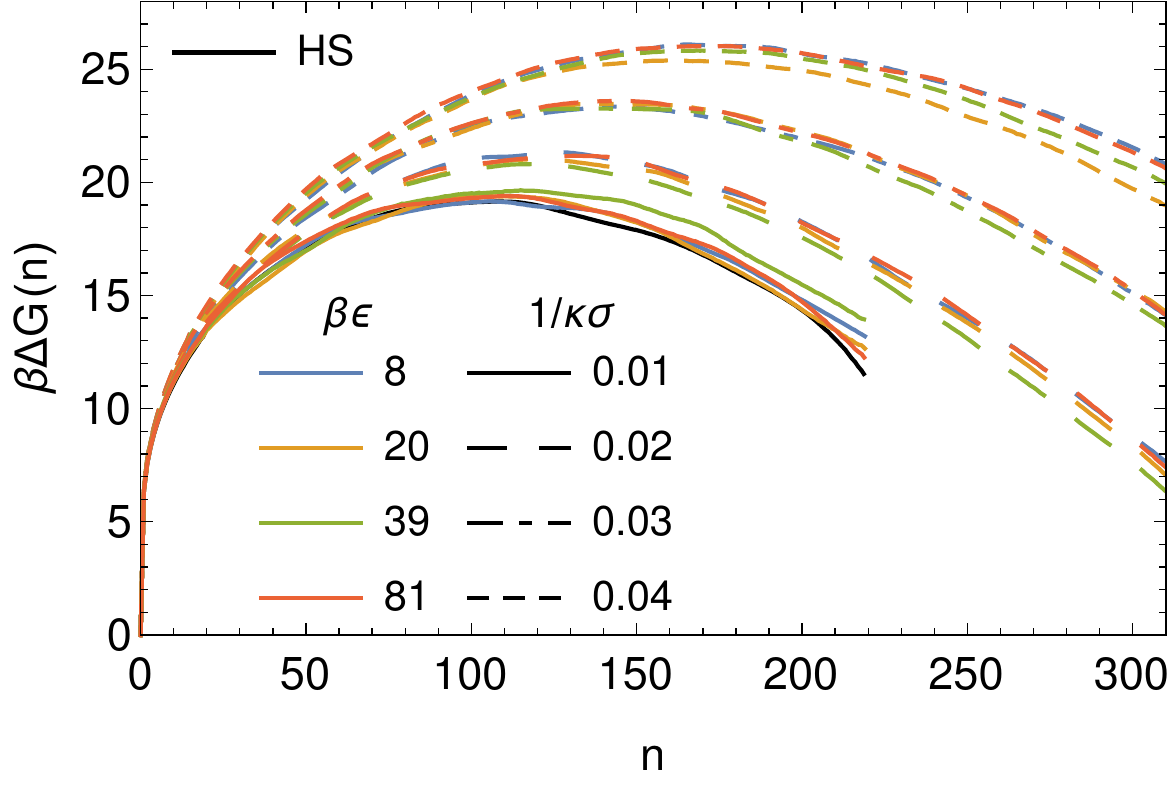}
    \caption{\label{fig:fcceffective} Nucleation barriers %of the FCC crystal
    of hard-core Yukawa particles with contact value $\beta\epsilon$ (colors) and screening length $1/\kappa\sigma$ (dashing) for a packing fraction of the supersaturated fluid of $\eta^*=1.088\eta_F$. The corresponding hard-sphere barrier is shown by a black solid line. For more information see Tab. \ref{tab:fccinfoeffective}.}
\end{figure}

\begin{table*}[t]
\caption{\label{tab:fccinfoeffective} Interfacial free energy  $\beta\gamma\sigma^2$, critical nucleus size $n^*$, and barrier height $\beta\Delta G^*$ obtained from fitting to the nucleation barriers of Fig. \ref{fig:fcceffective}. The third-fifth column give the cutoff radius $r_c/\sigma$, packing fraction $\eta^*$ of the supersaturated fluid, and pressure $\beta P\sigma^3$ used for the simulations, respectively. The sixth and seventh column give the density $\rho_s\sigma^3$ of the solid phase and supersaturation $\beta|\Delta\mu|$ used for fitting the barrier. The error in $\beta\Delta G^*$ is no more than 1.}
\begin{ruledtabular}
\begin{tabular}{ccccccccccc}
$\beta\epsilon$ & $1/\kappa\sigma$ & $r_c/\sigma$ & $\eta^*$ & $\beta P\sigma^3$ & $\rho_s \sigma^3$ & $\beta |\Delta\mu|$ & $\beta\gamma\sigma^2$ & $\beta\gamma/\rho_s^{2/3}$ & $n^*$ & $\beta\Delta G^*$ \\ \hline
\multicolumn{2}{c}{hard spheres}    & 1.40    & 0.5352    & 17.0              & 1.136             & 0.536               & 0.58  &  0.53     & 101   & 19.1                 \\ \hline
8  & 0.01 & 1.43    & 0.4985    & 16.0              & 1.053             & 0.531               & 0.53  &  0.52     & 103   & 19.2                 \\
20 & 0.01 & 1.45    & 0.4853    & 15.6              & 1.026             & 0.536               & 0.55  &  0.55     & 102    & 19.2                 \\
39 & 0.01 & 1.46    & 0.4757    & 15.2              & 1.005             & 0.530               & 0.53  &  0.53     & 108   & 19.6                 \\
81 & 0.01 & 1.47    & 0.4662    & 15.0              & 0.986             & 0.540               & 0.52  &  0.53     & 102    & 19.4                 \\ \hline
8  & 0.02 & 1.46    & 0.4699    & 15.4              & 0.983             & 0.494               & 0.57  &  0.57     & 116   & 21.2                 \\
20 & 0.02 & 1.49    & 0.4461    & 14.6              & 0.934             & 0.499               & 0.52  &  0.55     & 116   & 20.9                 \\
39 & 0.02 & 1.51    & 0.4298    & 14.1              & 0.901             & 0.502               & 0.52  &  0.56     & 114   & 20.7                 \\
81 & 0.02 & 1.53    & 0.4131    & 13.5              & 0.866             & 0.495               & 0.51  &  0.56     & 118   & 21.1                 \\ \hline
8  & 0.03 & 1.49    & 0.4477    & 15.1              & 0.927             & 0.452               & 0.52  &  0.55     & 142   & 23.2                 \\
20 & 0.03 & 1.52    & 0.4150    & 14.0              & 0.860             & 0.462               & 0.50  &  0.55     & 142   & 23.3                 \\
39 & 0.03 & 1.56    & 0.3932    & 13.2              & 0.815             & 0.458               & 0.49  &  0.57     & 137   & 23.2                 \\
81 & 0.03 & 1.58    & 0.3706    & 12.4              & 0.769             & 0.460               & 0.47  &  0.55     & 141   & 23.4                 \\ \hline
8  & 0.04 & 1.50    & 0.4304    & 15.1              & 0.882             & 0.411               & 0.50  &  0.54     & 172   & 25.8                 \\
20 & 0.04 & 1.56    & 0.3902    & 13.7              & 0.801             & 0.430               & 0.48  &  0.55     & 162   & 25.3                 \\
39 & 0.04 & 1.59    & 0.3630    & 12.6              & 0.746             & 0.424               & 0.45  &  0.54     & 168   & 25.7                 \\
81 & 0.04 & 1.63    & 0.3366    & 11.6              & 0.692             & 0.425               & 0.42  &  0.54     & 170   & 25.9                 \\
\end{tabular}
\end{ruledtabular}
\end{table*}

We now turn our attention to the nucleation behavior of these nearly-hard particles.  To study the crystal nucleation, we use MC simulations in the $NPT$-ensemble combined with umbrella sampling and measure the nucleation free-energy barriers.  Following Refs. \onlinecite{filion2010crystal,filion2011simulation}, we bias the simulation using the number of particles in the nucleus measured via the Ten Wolde order parameter \cite{tenwolde1996simulation} (see SM) with cutoff values $d_c=0.7$ and $\xi_c=6$. We use a fixed radial cutoff $r_c$ to determine the nearest neighbors which is chosen to be approximately the position of the first minimum of the radial distribution function for each state point. 

We compute the nucleation barriers for the various hard-core Yukawa systems at a fixed effective packing fraction of the supersaturated fluid of $\eta^*=1.088\eta_F$, which  corresponds to hard spheres at a pressure of $\beta P\sigma^3=17.0$. The resulting nucleation barriers are shown in Fig. \ref{fig:fcceffective}.  For each system we use $N=10976$, and Tab. \ref{tab:fccinfoeffective} contains all other necessary information per system. 
Additionally, Tab. \ref{tab:fccinfoeffective} gives the interfacial free energy, critical nucleus size, and barrier height obtained from fitting the nucleation barriers (see SM). Based on our estimates, the error in the barrier height is no more than $1k_BT$.
Looking at Fig. \ref{fig:fcceffective} and Tab. \ref{tab:fccinfoeffective}, we see that, similar to $\eta_M^\text{eff}$ in Fig. \ref{fig:potentials}b), the nucleation barriers are grouped together according to $1/\kappa\sigma$. From this we conclude that the nucleation barrier is mostly dependent on the screening length and only weakly dependent on the contact value. 
%This is not entirely surprising, as both $\eta_M^\text{eff}$ and $\beta|\Delta\mu|$ also depend mostly on $1/\kappa\sigma$.
Furthermore, we see that the systems with $1/\kappa\sigma=0.01$ map well to hard spheres, and that increasing $1/\kappa\sigma$, i.e. increasing the softness of the particles, increases the height of the nucleation barrier. This is not entirely surprising, as the supersaturation, see Tab. \ref{tab:fccinfoeffective}, decreases with increasing $1/\kappa\sigma$.

Next, we compute the nucleation rates for these systems. 
% As the nucleation rate is determined for the most part by the height of the nucleation barrier, and the barrier height in turn is mostly dependent on $1/\kappa\sigma$, we only compute the nucleation rate for one or two systems per $1/\kappa\sigma$.
To calculate the attachment rate, we select 10 independent configurations from the umbrella simulation of the window on top of the barrier and start 10 independent kinetic MC simulations from each configuration, see SM. The resulting attachment and nucleation rates in terms of the long-time diffusion coefficient $D_l$ are given in Tab. \ref{tab:fccrates}. The error in the attachment rate is approximately 10\%. Combining this with the $1k_BT$ error in the barrier height gives an error in the nucleation rate of approximately a factor 3. 
The nucleation rate found for the hard-spheres system agrees with the ones given in Ref. \onlinecite{filion2010crystal,auer2001prediction}. 
As expected from the increasing barrier height, we see that the nucleation rate decreases with increasing $1/\kappa\sigma$.  This is interesting, as this is in the opposite direction of the deviations between experiments and simulations for hard spheres. The ``softer'' particles studied in experiments have even lower rates than those predicted for purely hard spheres meaning that the discrepancy between experiments and simulations cannot be explained by softness due to charge.

\begin{table}[h]
\caption{\label{tab:fccrates} Nucleation rate $k$ in terms the long-time diffusion coefficient $D_l$ for a few of the systems of Tab. \ref{tab:fccinfoeffective}. The third and fourth column give the second derivative of the nucleation barrier and the attachment rate, both on top of the barrier, used for calculating $k$. The error in $k$ is approximately a factor 3.
}
\begin{ruledtabular}
\begin{tabular}{ccccc}
    $\beta\epsilon$ & $1/\kappa\sigma$ & $\beta\Delta G''(n^*)$ & $f_{n^*}/6D_l$ & $k\sigma^5/6D_l$\\[+0.1cm] \hline \\[-0.3cm]
    \multicolumn{2}{c}{hard spheres} & $-1.2\cdot10^{-3}$ & $2.1\cdot10^{3}$ & $1.5\cdot10^{-7}$\\
    8  & 0.01 & $-1.1\cdot10^{-3}$ & $2.5\cdot10^3$ & $1.5\cdot10^{-7}$\\
    81 & 0.01 & $-1.2\cdot10^{-3}$ & $6.6\cdot10^3$ & $3.1\cdot10^{-7}$\\
    8  & 0.02 & $-1.1\cdot10^{-3}$ & $2.8\cdot10^3$ & $2.1\cdot10^{-8}$\\
    8  & 0.03 & $-8.0\cdot10^{-4}$ & $3.0\cdot10^3$ & $2.5\cdot10^{-9}$\\
    8  & 0.04 & $-6.1\cdot10^{-4}$ & $3.3\cdot10^3$ & $1.7\cdot10^{-10}$\\
    81  & 0.04 & $-6.2\cdot10^{-4}$ & $4.9\cdot10^3$ & $1.7\cdot10^{-10}$\\
\end{tabular}
\end{ruledtabular}
\end{table}

\begin{figure}[h]
    \centering
    \includegraphics[width=\figwidthA]{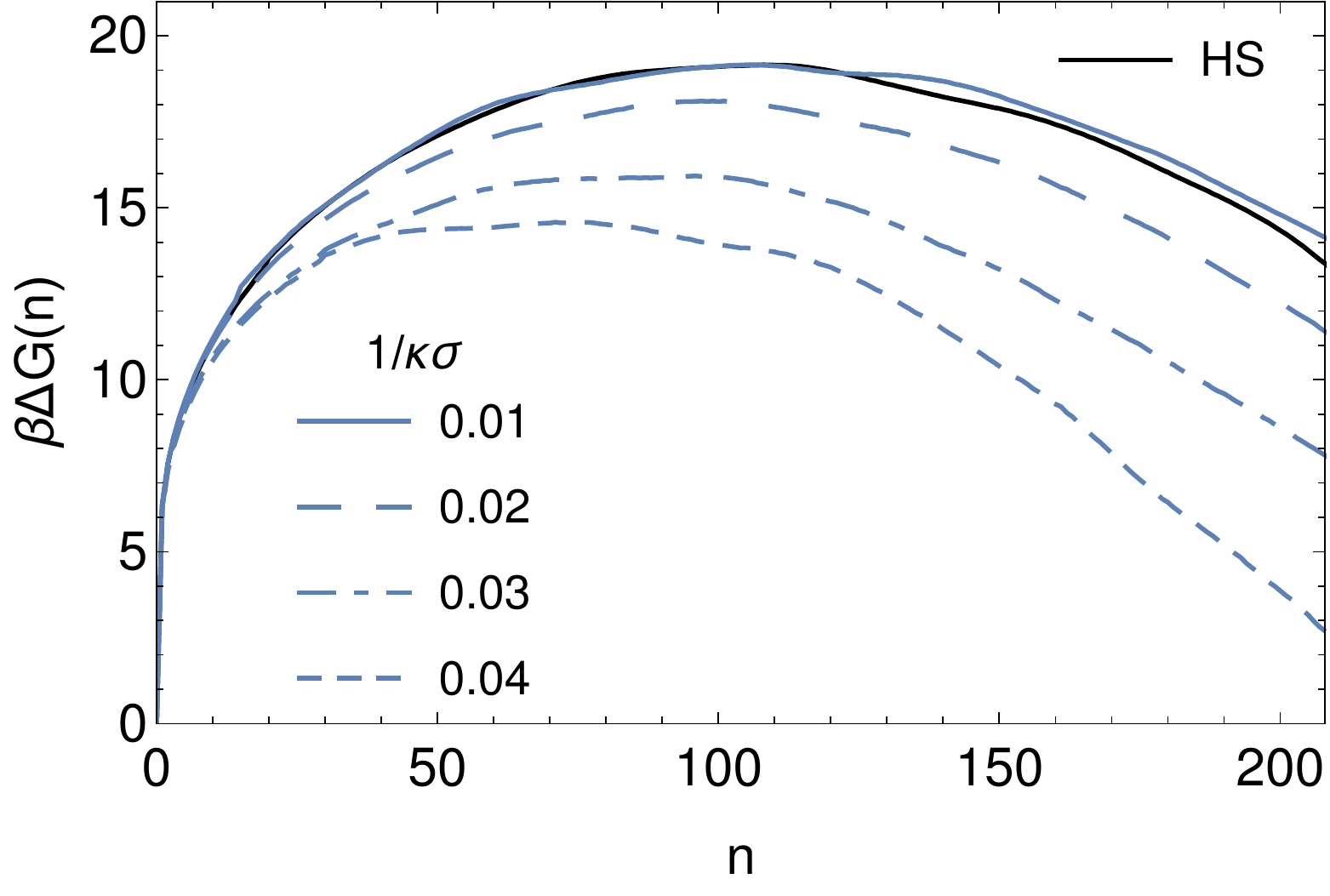}
    \caption{\label{fig:fccmu} Nucleation barriers of the FCC crystal of hard-core Yukawa particles with contact value $\beta\epsilon=8$ and screening length $1/\kappa\sigma$ (dashing) for a supersaturation of $\beta|\Delta\mu|\approx0.54$. The corresponding hard-sphere barrier is shown by black solid line. For more information see Tab. \ref{tab:fccinfomu}.}
\end{figure}

The question remains -- is there a different way of mapping to hard spheres that leads to a better agreement for the nucleation rates? One option would be to map the entire coexistence region, however this leads to even higher barriers. Another common practice is to compare nucleation rates at fixed $\beta |\Delta \mu|$. 
Here we also compute the nucleation barriers at $\beta|\Delta\mu|\approx0.54$ for $1/\kappa\sigma=0.02,0.03,$ and $0.04$ with $\beta\epsilon=8$. The necessary information for each system is given in Tab. \ref{tab:fccinfomu}, and Fig. \ref{fig:fccmu} shows the resulting nucleation barriers. Contrary to the mapping at equal effective packing fraction, we now see that increasing $1/\kappa\sigma$ decreases the height of the nucleation barrier, and again good agreement only occurs when $1/\kappa \sigma = 0.01$.

% \begin{figure*}[t]
% \begin{tabular}{lll}
%      a) & \hspace{0.5cm} & b)  \\[-0.3cm]
%      \includegraphics[width=\figwidthB]{barriers_fcc_mu54_2b.pdf} & & \includegraphics[width=\figwidthB]{barriers_fcc_mu58_2b.pdf}
% \end{tabular}
%     \caption[width=1\linewidth]{\label{fig:fccmu} Nucleation barriers of the FCC crystal of nearly-hard spheres with contact value $\beta\epsilon$ (colors) and screening length $1/\kappa\sigma$ (dashing) for a supersaturation of a)  $\beta|\Delta\mu|\approx0.54$ and b) $\beta|\Delta\mu|\approx0.58$. The corresponding hard-sphere barriers are given by the solid black line. For more information see Tab. \ref{tab:fccinfomu}. }
% \end{figure*}

\begin{table*}[t]
\caption{\label{tab:fccinfomu} Interfacial free energy $\beta\gamma\sigma^2$, critical nucleus size $n^*$, and barrier height $\beta\Delta G^*$ obtained from fitting the nucleation barriers of Fig. \ref{fig:fccmu}. 
The third-fifth column give the cutoff radius $r_c/\sigma$, packing fraction $\eta^*$ of the supersaturated fluid, and pressure $\beta P\sigma^3$ used for the simulations, respectively. The sixth and seventh column give the density $\rho_s\sigma^3$ of the solid phase and supersaturation $\beta|\Delta\mu|$ used for fitting the barrier. The error in $\beta\Delta G^*$ is no more than 1.}
\begin{ruledtabular}
\begin{tabular}{ccccccccccc}
$\beta\epsilon$ & $1/\kappa\sigma$ & $r_c/\sigma$ & $\eta^*$ & $\beta P\sigma^3$ & $\rho_s \sigma^3$ & $\beta |\Delta\mu|$ & $\beta\gamma\sigma^2$ & $\beta\gamma/\rho_s^{2/3}$ & $n^*$ & $\beta\Delta G^*$ \\ \hline
\multicolumn{2}{c}{hard spheres}    & 1.40    & 0.5353    & 17.0              & 1.136             & 0.536               & 0.58  &  0.53     & 101   & 19.1    \\ 
8  & 0.01 & 1.43    & 0.4985    & 16.0              & 1.053             & 0.531               & 0.53  &  0.52     & 103   & 19.2                 \\
%20 & 0.01 & 1.45    & 0.4850    & 15.6              & 1.026             & 0.533               & 0.55     & 0.54  & 102  & 19.2          \\
%39 & 0.01 & 1.46    & 0.4743    & 15.2              & 1.005             & 0.533               & 0.53     & 0.53  & 108  & 19.6          \\
%81 & 0.01 & 1.47    & 0.4657    & 15.0              & 0.986             & 0.541               & 0.52     & 0.53  & 102  & 19.4          \\ 
8  & 0.02 & 1.46    & 0.4724    & 15.8              & 0.988             & 0.532               & 0.51     & 0.51  & 93   & 18.0          \\
8  & 0.03 & 1.48    & 0.4538    & 16.1              & 0.938             & 0.541               & 0.45     & 0.47  & 81   & 16.0          \\
8  & 0.04 & 1.49    & 0.4399    & 16.7              & 0.900             & 0.541               & 0.41     & 0.44  & 69   & 14.8          \\ %\hline
% \multicolumn{2}{c}{hard spheres}     & 1.40    & 0.5385    & 17.5              & 1.142             & 0.585               & 0.57   & 0.52  & 75   & 16.5          \\
% 8  & 0.01 & 1.43    & 0.5039    & 16.6              & 1.061             & 0.591               & 0.49   & 0.47  & 80   & 15.9          \\
% 20 & 0.01 & 1.44    & 0.4884    & 16.1              & 1.032             & 0.587               & 0.48   & 0.47  & 81   & 15.8          \\
% 39 & 0.01 & 1.45    & 0.4772    & 15.7              & 1.011             & 0.583               & 0.51   & 0.51  & 80   & 16.7          \\
% 81 & 0.01 & 1.46    & 0.4681    & 15.4              & 0.991             & 0.584               & 0.49   & 0.49  & 84   & 16.3          \\ 
% 8  & 0.02 & 1.46    & 0.4753    & 16.3              & 0.994             & 0.580               & 0.47   & 0.48  & 75   & 15.7          \\
% 8  & 0.03 & 1.48    & 0.4567    & 16.6              & 0.943             & 0.585               & 0.43   & 0.45  & 63   & 14.4          \\
%8  & 0.04 & 1.49    & 0.4430    & 17.3              & 0.906             & 0.586               &        &           &        &   13?           \\ 
\end{tabular}
\end{ruledtabular}
\end{table*}

%%%%%%%%%%%%%%%%%%%%%%%%%%%%%%%%%%%%%%%%%%%%%%%%%%%%%
%%%%%%%%%%%%%%%%%%%%%%%%%%%%%%%%%%%%%%%%%%%%%%%%%%%%%

In conclusion, we demonstrated that for highly-screened particles the phase behavior depends nearly completely on the screening length, and not the charge.  Specifically, when the freezing densities of charged particles are mapped onto each other, the effective systems have the same melting densities, and the same nucleation barriers and rates.  Interestingly, the effective phase behavior depends extremely sensitively on the screening length, playing a measurable role even for screening lengths as low as $1/\kappa \sigma \sim 0.02$. This begs the question:  when can experimental systems be considered hard spheres?  The results here suggest that the answer is when the effective melting density matches that of hard spheres.

We would like to thank Frank Smallenburg, Michiel Hermes, and Alfons van Blaaderen for many useful discussions. L.F. and M.d.J. acknowledge funding from the Vidi research program with project number VI.VIDI.192.102 which is financed by the Dutch Research Council (NWO).

%%%%%%%%%%%%%%%%%%%%%%%%%%%%%%%%%%%%%%%%%%%%%%%%%%%%%
%%%%%%%%%%%%%%%%%%%%%%%%%%%%%%%%%%%%%%%%%%%%%%%%%%%%%

\clearpage

\bibliography{paper}% Produces the bibliography via BibTeX.

\end{document}

\preprint{APS/123-QED}

\title{Supplemental Material for ``Crystal nucleation of hard-core Yukawa particles''}% Force line breaks with \\

\author{Marjolein de Jager}
\author{Laura Filion}
\affiliation{Soft Condensed Matter, Debye Institute of Nanomaterials Science, Utrecht University, Utrecht, Netherlands}%Lines break automatically or can be forced with \\

% \collaboration{CLEO Collaboration}%\noaffiliation

\maketitle

\onecolumngrid  %for single column paper

%\tableofcontents

%\setcounter{figure}{0}
%\setcounter{table}{0}
\renewcommand{\thetable}{S\arabic{table}}
\renewcommand{\thefigure}{S\arabic{figure}}
\renewcommand{\theequation}{S\arabic{equation}}

%%%%%%%%%%%%%%%%%%%%%%%%%%%%%%%%%%%%%%%%%%%%%%%%%%%%%%%%%%%%%%%
%%%%%%%%%%%%%%%%%%%%%%%%%%%%%%%%%%%%%%%%%%%%%%%%%%%%%%%%%%%%%%%

\section{Coexistence densities}
In the main paper, we report the effective melting packing fractions as a function of $1/\kappa\sigma$ and $\beta\epsilon$ for the systems studied. Here, in  Table \ref{tab:fccfreezemelt}, we give additional information on the specific coexistence densities and resulting effective hard-sphere diameter $\sigma_\text{eff} /\sigma =  (\eta^\text{HS}_F / \eta_F)^{1/3}$ of these systems. The coexistence densities agree well with the phase diagrams given in Ref. \onlinecite{hynninen2003phase}. 
%Additionally, Tab. \ref{tab:fccfreezemelt} gives the resulting effective hard-sphere diameter $\sigma_\text{eff} /\sigma =  (\eta^\text{HS}_F / \eta_F)^{1/3}$. % used for scaling the radial distribution function and structure factor. 
As mentioned in the main paper, for $\beta\epsilon>1.5$, the effective melting packing fraction $\eta^\text{eff}_M$ is nearly unaffected by changing $\beta\epsilon$. To better understand this observation, we take a closer look at the interaction potential as a function of $r/\sigma_\text{eff}$ around $r=\sigma_\text{eff}$, see Fig. \ref{fig:potentialsscaled}. From this figure it immediately becomes apparent that the systems are neatly grouped together according to their screening length. Hence, it should not come as a surprise that systems with practically identical effective potentials also have practically identical effective melting packing fractions. Note that this mapping is only possible {\it after} determining the coexistence densities, and hence does not provide a way in which to determine in advance which potentials will map well onto hard spheres.

%To better understand the effect of scaling a system with this $\sigma_\text{eff}$, Fig. \ref{fig:potentialsscaled} takes a closer look at the interaction potential as a function of $r/\sigma_\text{eff}$ around $r=\sigma_\text{eff}$. One thing that becomes immediately apparent from this figure, is that the systems group together according to the screening length. 

\begin{table}[b]
\caption{\label{tab:fccfreezemelt} Freezing and melting packing fractions, $\eta_F$ and $\eta_M$, of various systems of nearly-hard spheres with contact value $\beta\epsilon$ and screening length $1/\kappa\sigma$. The last two columns of both tables give the effective hard-sphere melting packing fraction $\eta^\text{eff}_M =  (\eta^\text{HS}_F / \eta_F) \, \eta_M$, with $\eta^\text{HS}_F=0.492(1)$, and effective hard-spheres diameter $\sigma_\text{eff} /\sigma =  (\eta^\text{HS}_F / \eta_F)^{1/3}$. 
Note that $\eta^\text{HS}_M=0.543(1)$ \cite{frenkelbook,polson2000finite}, and the error in $\eta_F$ and $\eta_M$ is approximately 0.001.
To compare, the last two rows of the table on the right give the values for the WCA system of Ref. \onlinecite{filion2011simulation} and experimental PMMA particles of Pusey and van Megen \cite{pusey1986phase}.
}
\begin{minipage}{0.47\textwidth}
\begin{ruledtabular}
\begin{tabular}{cccccc}
$\beta\epsilon$ & $1/\kappa\sigma$ & $\eta_F$ & $\eta_M$ & $\eta^\text{eff}_M$ & $\sigma_\text{eff} /\sigma$ \\
\hline
8  & 0.01 & 0.458    & 0.505    & 0.542 & 1.024 \\
20 & 0.01 & 0.446    & 0.491    & 0.542 & 1.033 \\
39 & 0.01 & 0.437    & 0.482    & 0.542 & 1.040 \\ 
81 & 0.01 & 0.429    & 0.472    & 0.541 & 1.047 \\ \hline
8  & 0.02 & 0.432    & 0.473    & 0.538 & 1.044 \\
20 & 0.02 & 0.410    & 0.449    & 0.538 & 1.063 \\
39 & 0.02 & 0.395    & 0.432    & 0.538 & 1.076 \\
81 & 0.02 & 0.380    & 0.415    & 0.538 & 1.090 \\ \hline
8  & 0.03 & 0.412    & 0.447    & 0.534 & 1.061 \\
20 & 0.03 & 0.381    & 0.414    & 0.534 & 1.089 \\
39 & 0.03 & 0.361    & 0.393    & 0.535 & 1.108 \\
81 & 0.03 & 0.341    & 0.371    & 0.535 & 1.130  \\ \hline
8  & 0.04 & 0.396    & 0.426    & 0.530 & 1.075  \\
20 & 0.04 & 0.358    & 0.386    & 0.530 & 1.111  \\
39 & 0.04 & 0.334    & 0.360    & 0.531 & 1.138  \\
81 & 0.04 & 0.310    & 0.334    & 0.531 & 1.167  \\ 
\end{tabular}
\end{ruledtabular}
\end{minipage} \hfill
\begin{minipage}{0.47\textwidth}
\begin{ruledtabular}
\begin{tabular}{cccccc}
$\beta\epsilon$ & $1/\kappa\sigma$ & $\eta_F$ & $\eta_M$ & $\eta^\text{eff}_M$ & $\sigma_\text{eff} /\sigma$ \\
\hline
0.5& 0.04 & 0.476    & 0.523    & 0.541 & 1.011  \\
1  & 0.04 & 0.472    & 0.512    & 0.534 & 1.014  \\
1.5& 0.04 & 0.463    & 0.500    & 0.531 & 1.021  \\
2  & 0.04 & 0.453    & 0.489    & 0.530 & 1.028  \\
3  & 0.04 & 0.439    & 0.472    & 0.529 & 1.039  \\
4  & 0.04 & 0.426    & 0.458    & 0.529 & 1.049  \\
5  & 0.04 & 0.415    & 0.447    & 0.529 & 1.058  \\
6  & 0.04 & 0.408    & 0.439    & 0.529 & 1.065  \\
7  & 0.04 & 0.403    & 0.434    & 0.530 & 1.069  \\
8  & 0.04 & 0.396    & 0.426    & 0.530 & 1.075  \\ \hline
\multicolumn{2}{c}{WCA ($\beta\epsilon=40$)} & 0.373  & 0.411 &  0.542 &  1.097  \\ 
\multicolumn{2}{c}{Pusey \& v. Megen} & 0.407  & 0.442 &  0.534\footnote{Note that Pusey and van Megen used $\eta^\text{HS}_F=0.494$, whereas we use $\eta^\text{HS}_F=0.492$. Here we have corrected these values for this difference.} &  1.065$^{\text{a}}$  \\
\end{tabular}
\end{ruledtabular}
\vspace{0.327cm} %\vspace{4.185cm}
\end{minipage}
\end{table}

\newcommand{\figwidthA}{0.5\linewidth}
 \begin{figure}[t]
     \centering
     \includegraphics[width=\figwidthA]{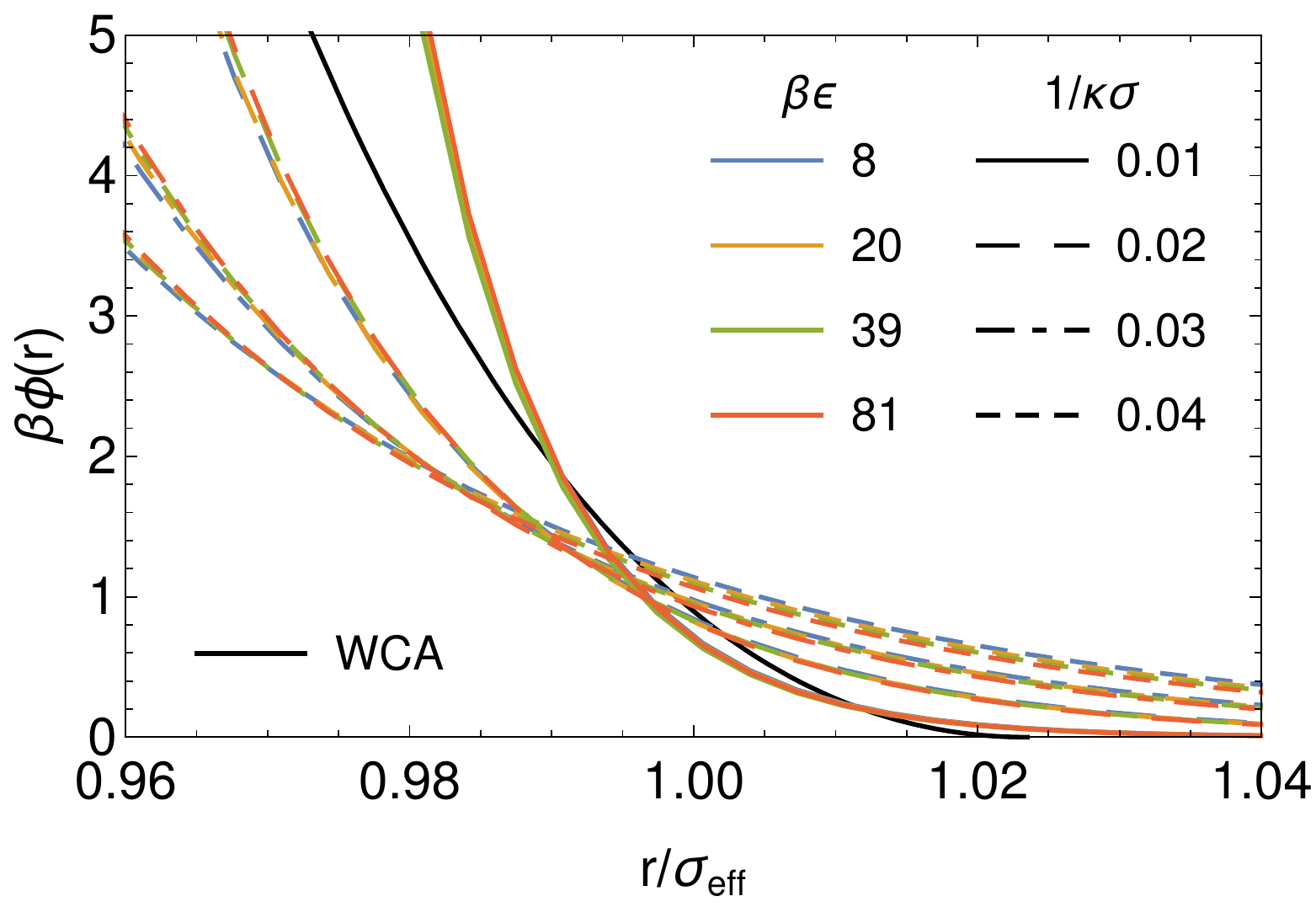}
     \caption{\label{fig:potentialsscaled} Interaction potential around at the effective hard-sphere diameter $\sigma_\text{eff}$ for hard-core Yukawa particles with contact value $\beta\epsilon$ (colors) and screening length $1/\kappa\sigma$ (dashing), plotted in units of $\sigma_\text{eff}$ (see Tab. \ref{tab:fccfreezemelt}). The solid black line shows the WCA potential with $\beta\epsilon=40$.}
 \end{figure}

%%%%%%%%%%%%%%%%%%%%%%%%%%%%%%%%%%%%%%%%%%%%%%%%%%%%%%%%%%%%%%%
%%%%%%%%%%%%%%%%%%%%%%%%%%%%%%%%%%%%%%%%%%%%%%%%%%%%%%%%%%%%%%%

\section{Fluid structure at melting}
In the main paper, we compare the structure of various fluids of (nearly-)hard particles at the freezing packing fraction by scaling with the effective hard-sphere diameter $\sigma_\text{eff}$. Here, to further demonstrate the accurateness of this mapping, we compare the structure of these fluids at the melting packing fraction.
Figure \ref{fig:structuremelt} shows the radial distribution function $g(r)$ and the structure factor $S(q)$ for these fluids. 
Again we see that scaling with $\sigma_\text{eff}$ results in an excellent mapping of the $g(r)$ and $S(q)$, and that the most significant differences, i.e. the very slight broadening of the first peak of the $g(r)$ and the mild shrinking of the higher order peaks of the $S(q)$, increase with increasing softness ($1/\kappa \sigma$).

\newcommand{\figwidthB}{0.45\linewidth}
\begin{figure*}[t!]
\begin{tabular}{lll}
     a) & \hspace{0.5cm} & b)  \\[-0.4cm]
     \includegraphics[width=\figwidthB]{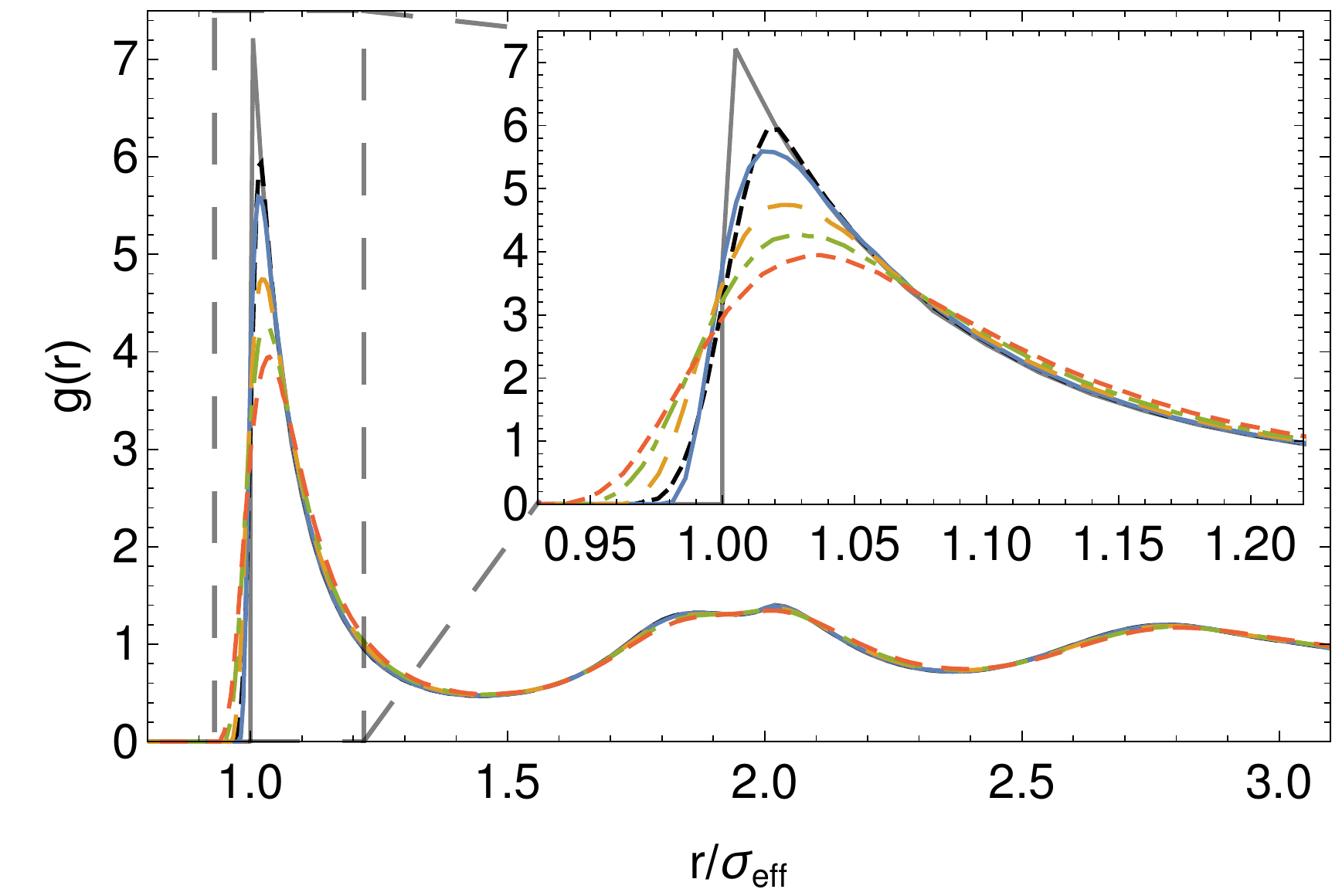} & & \includegraphics[width=\figwidthB]{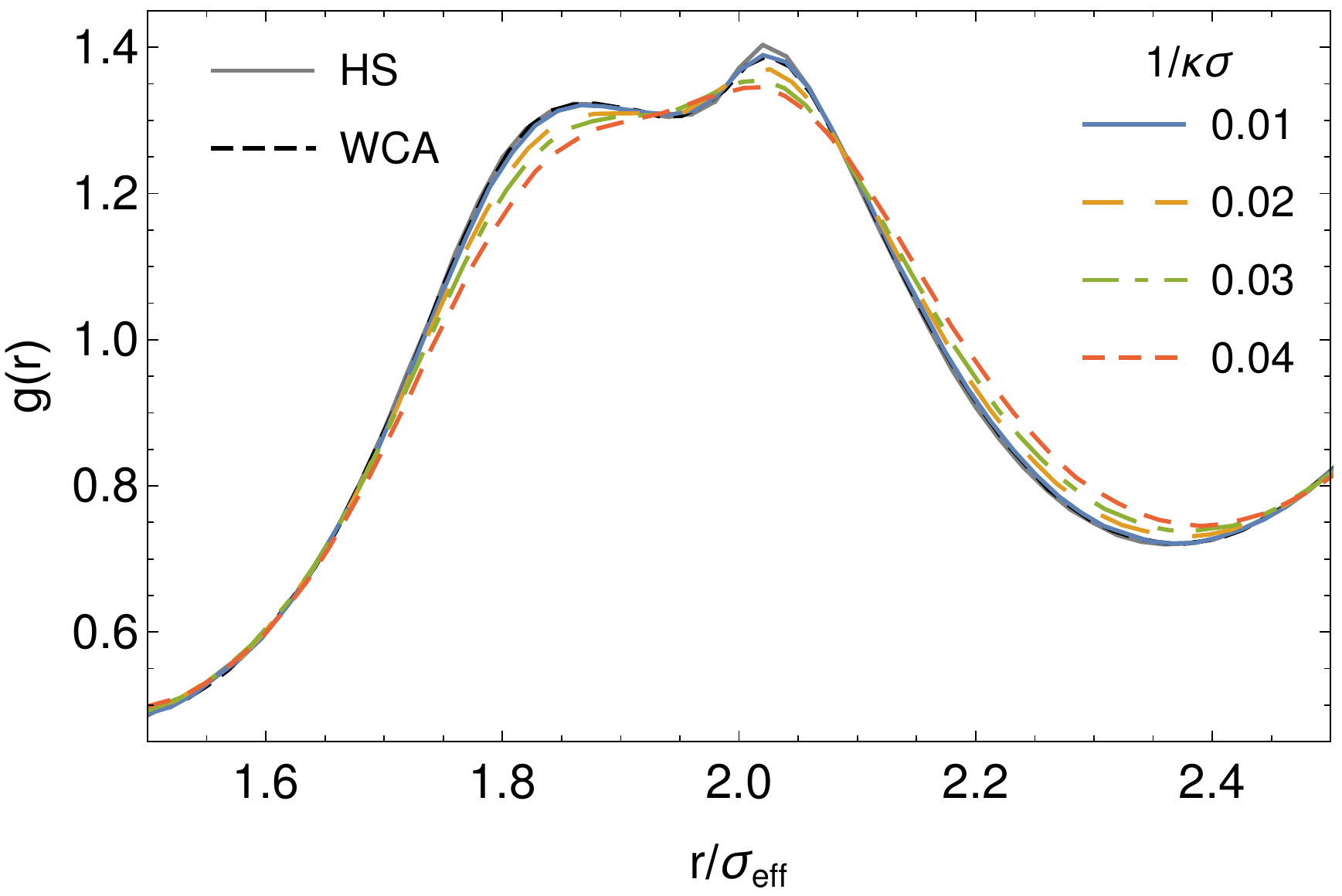} \\
     c) & \hspace{0.5cm} & d)  \\[-0.4cm]
     \includegraphics[width=\figwidthB]{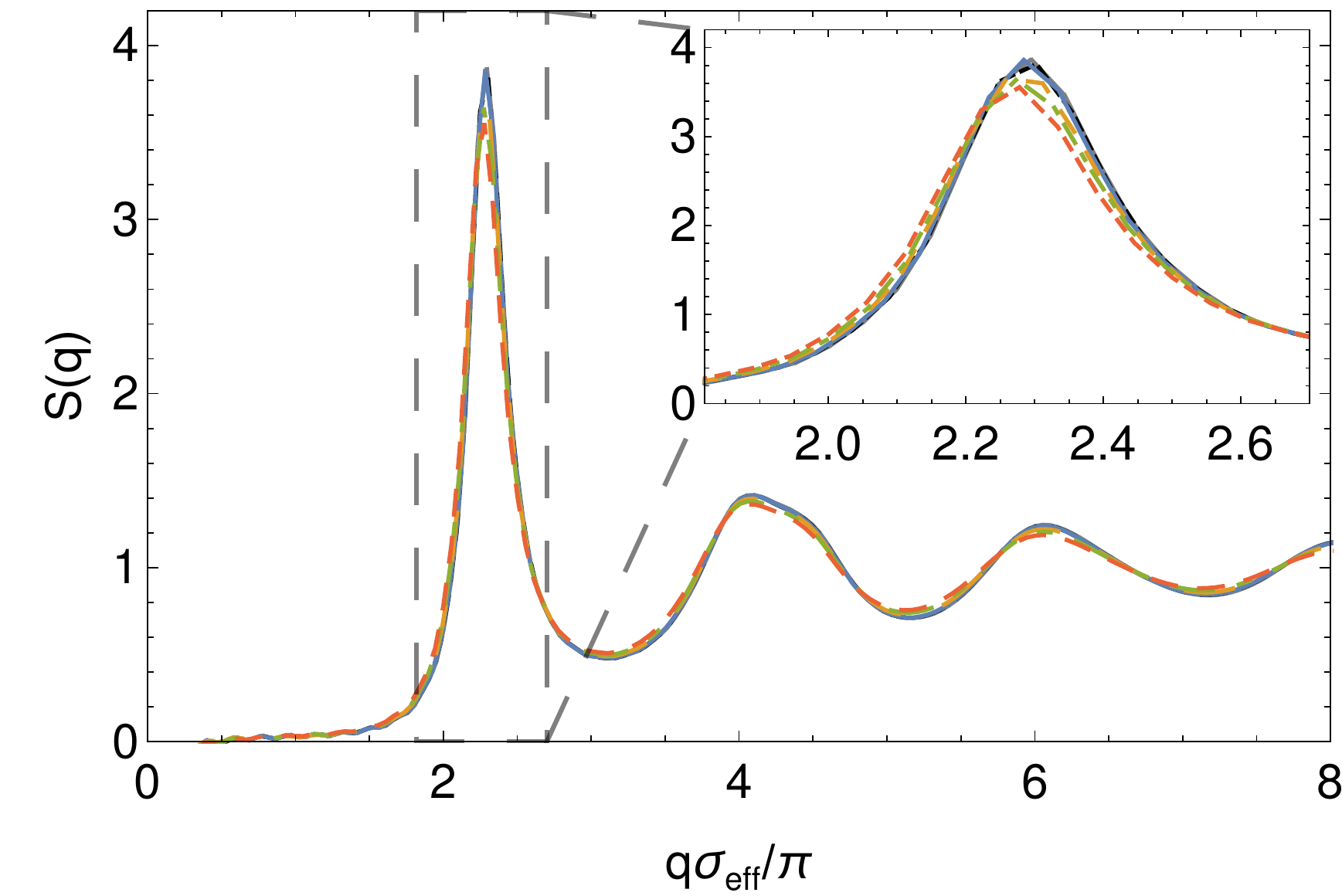} & & \includegraphics[width=\figwidthB]{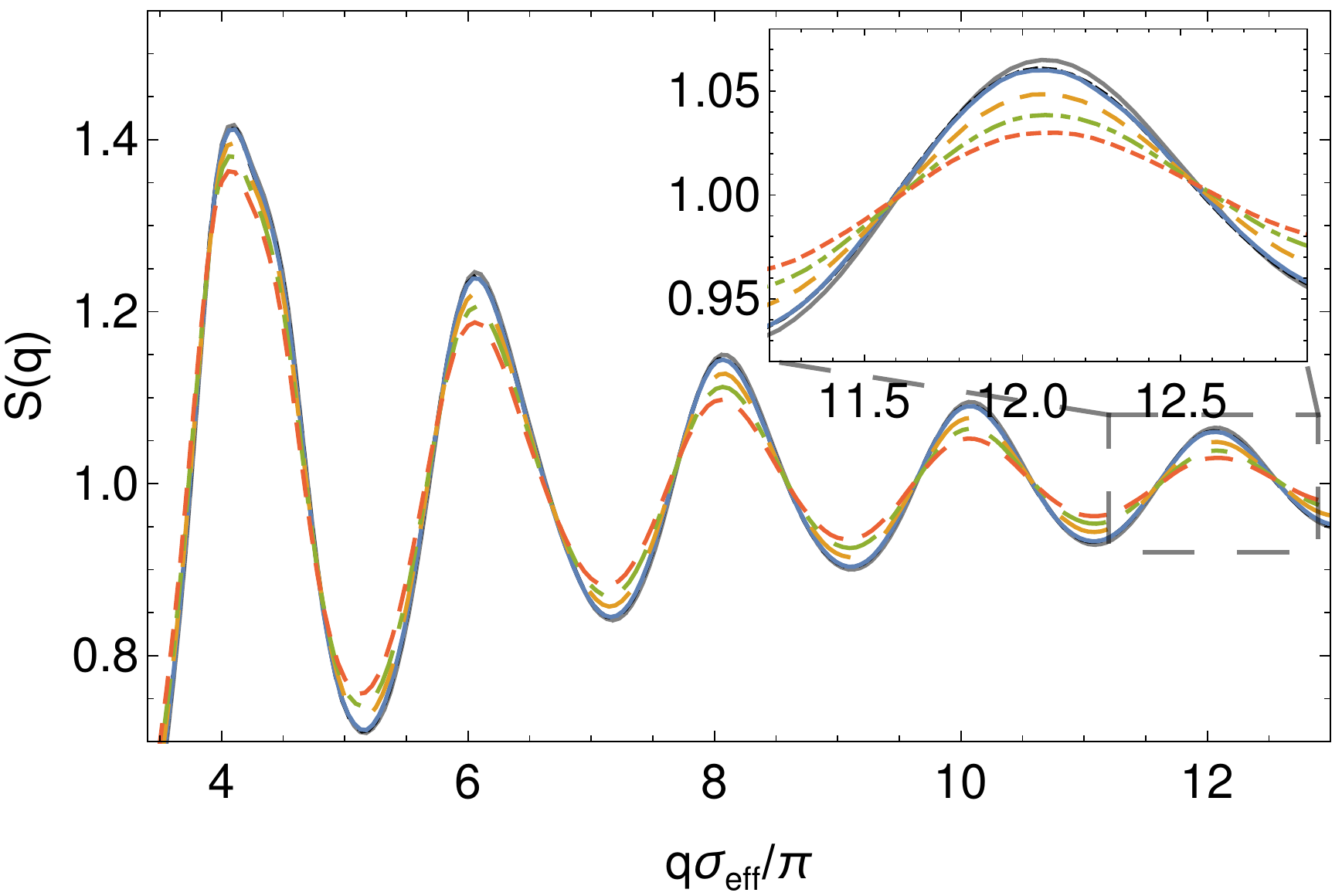}
\end{tabular}
    \caption[width=1\linewidth]{\label{fig:structuremelt} a,b) The radial distribution function $g(r)$ and c,d) structure factor $S(q)$ for a fluid of hard spheres (gray, solid), WCA particles with $\beta\epsilon=40$ (black, dashed), and hard-core Yukawa particles with $\beta\epsilon=39$ and varying screening length $1/\kappa\sigma$ (colors, dashing). All fluids were studied at the melting packing fraction and the horizontal axes are scaled by the effective hard-sphere diameter $\sigma_\text{eff}$, see Tab \ref{tab:fccfreezemelt}.
    The inset of a) zooms in on the first peak of the $g(r)$, and b) shows the second peak. The inset of c) zooms in on the first peak of the $S(q)$, and d) shows the second to sixth peak with the inset zooming in on the sixth peak.
    }
\end{figure*}

%%%%%%%%%%%%%%%%%%%%%%%%%%%%%%%%%%%%%%%%%%%%%%%%%%%%%%%%%%%%%%%
%%%%%%%%%%%%%%%%%%%%%%%%%%%%%%%%%%%%%%%%%%%%%%%%%%%%%%%%%%%%%%%

\section{Nucleation barriers and rates}

%%%%%%%%%%%%%%%%%%%%%%%%%%%%%%%%%%%%%%%%%%%%%%%%%%%%%%%%%%%%%%%

\subsection{Order parameter}
We use the number of particles of the solid nucleus $n$ as an order parameter for studying crystal nucleation \cite{tenwolde1996simulation,auer2001prediction,auer2004numerical}. 
One of the most commonly used techniques for differentiating between liquid and solid on a single-particle level is the one introduced by Ten Wolde \textit{et al.} \cite{tenwolde1996simulation}. 
%One of the most commonly used techniques for differentiating between liquid-like and solid-like particles is the one introduced by Ten Wolde \textit{et al.} \cite{tenwolde1996simulation}. 
This technique uses the bond orientational order parameters $q_{lm}$ to determine the number of solid-like connections $\xi$ of each particle and labels a particle as solid when $\xi\geq \xi_c$, with $\xi_c$ a cutoff value.
%Considering that we study nucleation of FCC and BCC crystals, we use the six-fold bond orientation order parameters
These bond-orientational order parameters are given by
\begin{equation}
\label{eq:q6m}
    q_{lm}(i) = \frac{1}{N_b(i)} \sum_{j\in\mathcal{N}_b(i)} Y_{lm}\left( \theta_{ij},\phi_{ij} \right),
\end{equation}
where $N_b(i)$ is the number of neighbors of particle $i$, $\,\mathcal{N}_b(i)$ is the set of neighbors of $i$, $\,Y_{lm}\left( \theta,\phi \right)$ are the spherical harmonics with $m\in[-l,l]$, and $\theta_{ij}$ and $\phi_{ij}$ are the polar and azimuthal angles of $\mathbf{r}_{ij}=\mathbf{r}(j)-\mathbf{r}(i)$, and $\mathbf{r}(i)$ is the position of particle $i$.
The number of solid-like connections of particle $i$ is then  determined via
\begin{equation}
    \xi(i) = \sum_{j\in\mathcal{N}_b} H\left(d_l(i,j) -d_c \right),
\end{equation}
where $H$ is the Heaviside step function, $d_c$ is the dot-product cutoff, and $d_l(i,j)$ is the dot product given by
%\begin{equation}
%    d_6(i,j) = \frac{ \sum_{m=-6}^{m=6} q_{6m}(i) q_{6m}^*(j) }{ \left(\sum_{m=-6}^{m=6} |q_{6m}(i)|^2 \right)^{1/2} \left(\sum_{m=-6}^{m=6} |q_{6m}(j)|^2 \right)^{1/2} },
%\end{equation}
\begin{equation}
    d_l(i,j) = \frac{ \sum_{m=-l}^{m=l} q_{lm}(i) q_{lm}^*(j) }{ \sqrt{ \left(\sum_{m=-l}^{m=l} |q_{lm}(i)|^2 \right) \left(\sum_{m=-l}^{m=l} |q_{lm}(j)|^2 \right) } },
\end{equation}
with $^*$ indicating the complex conjugate.
The neighbors of particle $i$ are defined as all particles $j$ with $|\mathbf{r}_{ij}|<r_c$, and a cluster contains all solid particles that have a solid neighbor in the same cluster. For all simulations we use $l=6$ and the cutoff values $d_c=0.7$ and  $\xi_c=6$. The nearest neighbour cutoff $r_c$  is chosen to be approximately the position of the first minimum of the radial distribution function for each state point.

%%%%%%%%%%%%%%%%%%%%%%%%%%%%%%%%%%%%%%%%%%%%%%%%%%%%%%%%%%%%%%%

\subsection{Umbrella sampling}
The Gibbs free-energy barrier associated with crystal nucleation can be determined via \cite{auer2001prediction,auer2004numerical}
\begin{equation}
    \beta \Delta G(n) = \text{constant} - \ln{P(n)},
\end{equation}
where $P(n)$ is the probability of observing a cluster of size $n$.
To measure $P(n)$ we use MC simulations in the $NPT$-ensemble combined with umbrella sampling \cite{frenkelbook,torrie1977nonphysical,van1992computer}. This technique  adds a biasing potential to the interaction potential of the system that drives the simulation to a preferred region of configuration space by artificially making a specific part of phase space more probable. The typical biasing potential used for computing nucleation barriers is given by \cite{tenwolde1996numerical,tenwolde1996simulation,filion2010crystal}
\begin{equation}
    \beta U_\text{bias}\left(n(\mathbf{r}^N) \right) = \frac{\lambda}{2} \left( n(\mathbf{r}^N) -n_c  \right)^2,
\end{equation}
where $\lambda$ is a coupling parameter, $n(\mathbf{r}^N)$ is the size of the largest cluster present in the configuration $\mathbf{r}^N$, and $n_c$ is the target cluster size. 

Thus, by choosing a suitable value for $\lambda$, we can force $n(\mathbf{r}^N)$ to fluctuate around $n_c$. Then, by performing the simulation for multiple values of $n_c$, we can compute different ``windows'' of the barrier. 
To compute a window, we measure the biased probability distribution $P_\text{bias}(n;n_c)$ for clusters of size $n$ and calculate the corresponding Gibbs free energy using
\begin{equation}
    \beta \Delta G(n;n_c) = \chi(n_c) - \ln{P_\text{bias}(n;n_c)} - \frac{\lambda}{2} \left(n-n_c \right)^2,
\end{equation}
where $\chi(n_c)$ is a constant shift needed to stitch the different windows together nicely \cite{auer2004numerical}. By ensuring that consecutive windows overlap, the shifts can be calculated.
For all nucleation barriers, we use $\lambda=0.02$, take $n_c$ with an interval of 10, and perform 4 independent runs for each window.
Note that for the first window, the barrier is determined without the use of a biasing potential and takes all present clusters into account. For the remainder of the barrier the presence of small clusters can be neglected \cite{auer2004numerical}.

\subsection{Fitting the nucleation barrier}
We compare our nucleation barriers to classical nucleation theory (CNT). In this simple theoretical model, it is assumed that crystal nucleation is controlled by the competition between the free-energy gain of the bulk crystal phase with respect to the fluid phase and the free-energy cost of making a fluid-crystal surface interface. More specifically, the Gibbs free-energy cost for making a spherical nucleus of radius $R$ is given by
\begin{equation}
\label{eq:cnt}
    \Delta G(R) = 4\pi\gamma R^2 - \frac{4}{3}\pi |\Delta\mu | \rho_s R^3,
\end{equation}
where $\gamma$ is the fluid-crystal interfacial free energy, $|\Delta\mu |$ is the difference in chemical potential between the crystal and fluid phases, and $\rho_s$ is the density of the crystal phase. The maximum height $\Delta G^*$ and critical radius $R^*$ of the nucleation barrier described by Eq. \eqref{eq:cnt} are given by
\begin{equation}
\label{eq:barrierheight}
    \Delta G^* = \frac{16\pi\gamma^3}{3 |\Delta\mu |^2 \rho_s^2}; \quad\quad\quad R^* = \frac{2\gamma}{|\Delta\mu | \rho_s}.
\end{equation}

The nucleation barrier can be fitted to CNT by realizing that the measured radius depends on the specific choice for $d_c$ and $\xi_c$. Assuming that the measured radius $R(d_c,\xi_c)$ differs from the unique radius associated with CNT with a constant shift $\alpha(d_c,\xi_c)$ \cite{filion2010crystal},  we obtain the equation
\begin{equation}
\label{eq:cntfit}
    \beta\Delta G(n; d_c, \xi_c)  = \delta + 4\pi\beta\gamma \left[ \left( \frac{3n(d_c,\xi_c)}{4\pi \rho_s} \right)^{1/3} - \alpha(d_c,\xi_c) \right]^2  -  \frac{4\pi}{3}\beta|\Delta\mu | \rho_s \left[ \left( \frac{3n(d_c,\xi_c)}{4\pi \rho_s} \right)^{1/3} - \alpha(d_c,\xi_c) \right]^3,
\end{equation}
where we used that the measured cluster size $n(d_c,\xi_c)$ can be related to the radius $R(d_c,\xi_c)$ by
\begin{equation}
    n(d_c,\xi_c) = \frac{4\pi R(d_c,\xi_c)^3 \rho_s}{3}.
\end{equation}
The constant shift $\delta$ is added to Eq. \eqref{eq:cntfit} for the reason that the nucleation barrier obtained via umbrella sampling is only expected to match CNT near the top of the barrier. 
Note that the specific choice for $d_c$ and $\xi_c$ does not affect the barrier height, only the width \cite{filion2010crystal}, and by using Eq. \eqref{eq:cntfit} to fit the barrier one can also obtain the interfacial free energy $\gamma$ independent of the choice for $d_c$ and $\xi_c$.

%%%%%%%%%%%%%%%%%%%%%%%%%%%%%%%%%%%%%%%%%%%%%%%%%%%%%%%%%%%%%%%

\subsection{Nucleation rate}
The nucleation rate per unit volume $k$ is related to the nucleation barrier by \cite{oxtoby1992homogeneous,auer2001prediction}
\begin{equation}
\label{eq:nuclrate}
    k = A e^{-\beta\Delta G^*},
\end{equation}
where $e^{-\beta\Delta G^*}$ is the probability of forming a critical nucleus, and $A$ is a kinetic factor, which provides a measure for the rate with which a critical nucleus grows. The latter can be approximated as \cite{oxtoby1992homogeneous,auer2001prediction,auer2004numerical}
\begin{equation}
    A \approx \rho f_{n^*} \sqrt{\frac{| \beta \Delta G''(n^*) |}{2\pi}},
\end{equation}
with $n^*$ the size of the critical nucleus, $\rho$ the density of the supersaturated fluid, $f_{n^*}$ the rate at which particles are attached to the critical nucleus, and $\Delta G''(n)$ the second derivative of the nucleation barrier.
The attachment rate is related to the mean squared deviation of the cluster size $\langle\Delta n^2(t)\rangle = \langle \left( n(t) - n^* \right)^2 \rangle$ at the top of the barrier by \cite{auer2001prediction,auer2004numerical}
\begin{equation}
    f_{n^*} = \frac{1}{2} \frac{\langle \Delta n^2(t)\rangle}{t}.
\end{equation}
%where $\Delta n^2(t) = \left( n(t) - n^* \right)^2$. 
To compute $\langle\Delta n^2(t)\rangle$, we start numerous kinetic Monte Carlo (KMC) simulations in the $NVT$-ensemble at the top of the barrier and measure $n(t)$. We then use the long-time behavior of $\langle\Delta n^2(t)\rangle$ to compute $f_{n^*}$ \cite{filion2010crystal}.
In the main paper, we report $k$ and $f_{n^*}$ in terms of the long-time diffusion coefficient $D_l$. To obtain $D_l$, we simply fit the long-time diffusion of the mean-squared displacement obtained from KMC simulations of the supersaturated fluid at the appropriate packing fraction.

%%%%%%%%%%%%%%%%%%%%%%%%%%%%%%%%%%%%%%%%%%%%%%%%%%%%%%%%%%%%%%%
%%%%%%%%%%%%%%%%%%%%%%%%%%%%%%%%%%%%%%%%%%%%%%%%%%%%%%%%%%%%%%%

\bibliography{paper}% Produces the bibliography via BibTeX.